\documentclass{amsart}
\usepackage{amsmath}
\usepackage{amsthm}
\usepackage{amsfonts}
\usepackage{amssymb}
\newtheorem{Theorem}{Theorem}[section]
\newtheorem{Proposition}[Theorem]{Proposition}
\newtheorem{Lemma}[Theorem]{Lemma}
\newtheorem{Corollary}[Theorem]{Corollary}
\theoremstyle{definition}
\newtheorem{Definition}{Definition}[section]
\numberwithin{equation}{section}

\newcommand{\Z}{{\mathbb Z}}
\newcommand{\R}{{\mathbb R}}

\newcommand{\N}{{\mathbb N}}

\begin{document}
\title[Finite propagation speed]{Finite propagation speed and
kernel estimates for Schr\"odinger operators}

\author{Christian Remling}

\address{Mathematics Department\\
University of Oklahoma\\
Norman, OK 73019-0315}
\email{cremling@math.ou.edu}
\urladdr{www.math.ou.edu/$\sim$cremling}
\date{\today}

\begin{abstract}
I point out finite propagation speed phenomena for discrete and continuous
Schr\"odinger operators and discuss various types of kernel estimates
from this point of view.
\end{abstract}
\maketitle
\section{Introduction}
In this note, I will point out finite propagation speed phenomena
associated with continuous Schr\"odinger operators
\begin{equation}
\label{soc}
(Hu)(x) = -\Delta u(x) + V(x)u(x)
\end{equation}
on $L_2(\R^d)$ and their discrete analogs
\begin{equation}
\label{sod}
(Hu)(n) = \sum_{|m-n|_1=1} u(m) + V(n)u(n) ,
\end{equation}
acting on $\ell_2(\Z^d)$. I believe that these observations lead to a very
transparent, non-technical and elegant treatment of various topics.

It is instructive to look at the one-dimensional half line problems for a moment.
So, consider the operators from \eqref{soc} and \eqref{sod}, acting on
$L_2(0,\infty)$ and $\ell_2(\N)$, respectively. Let $\rho$ be the standard
spectral measure; it can be obtained from the Weyl circle construction (see,
for example, \cite{CodLev}). Then, in the discrete case, the \textit{moments}
$\int \lambda^n \, d\rho(\lambda)$ are a very important object, and experience
has shown that a similar role is played by the function
\[
\phi(t) = \int \cos t\sqrt{\lambda}\, d\rho(\lambda) - 2\delta (t)
\]
in the continuous case.
(It can be shown that this formula, suitably interpreted, indeed defines
an absolutely continuous function. See, for example, \cite{RemdB}.)
In other words, there seems to be something special about the functions
$\lambda^n$ and $\cos t\sqrt{\lambda}$, respectively. As we will see in
Sect.\ 2, the unifying theme is \textit{finite propagation speed,} and this
in fact works in any dimension.

We will illustrate our basic observations
(see Lemmas \ref{Lbasicd}, \ref{Lbasicc} below) further by using them
to discuss the following two problems:

1.\ Let $f$ be a bounded smooth function
on the spectrum of $H$. Try to estimate $\langle \varphi_1, f(H) \varphi_2\rangle$
in terms of the separation of the supports of $\varphi_1$ and $\varphi_2$.

2.\ Suppose that $V$ is known on $B_R(0)=\{ x\in\R^d : |x|< R \}$ and $\varphi$ is supported
near zero. What can one then say about $\langle \varphi, f(H) \varphi \rangle$?

Both problems have been studied before by other methods, the first one
in fact quite extensively (see \cite{BGK} and the references cited therein).
The sample results we will prove in Sections 3, 4 below are quite similar
to what had been known before.
Actually, Corollary \ref{C5.2} below does seem to
improve results from \cite{BGK},
but that is definitely not my main point in this context.
Rather, what I'm trying to emphasize here is the realization that
finite propagation speed phenomena are at the heart of the matter,
and things become very transparent if this point of view is adopted.
This basic idea might be useful in other situations too.

A piece of closely related work is \cite{CGT}, where
finite propagation speed methods are used
to estimate kernels of functions of the Laplace-Beltrami operator on Riemannian
manifolds. In \cite{Sik}, these techniques are used to estimate heat kernels.
Finally, see \cite{GKT,Remuniv} for work on problem 2, and it may also be
interesting to take a look at \cite{LP} or \cite[Sect.\ XI.11]{RS3} for the use of
wave phenomena in scattering theory.

\medskip
\noindent\textit{Acknowledgments.} I thank Wilhelm Schlag for showing me the
proof of Lemma \ref{Lwe} and Adam Sikora for informing me of his work.
\section{Finite propagation speed}
We make the following \textbf{basic assumptions:} In the continuous case, we assume that
$H$ is essentially self-adjoint on $C_0^{\infty}(\R^d)$ and bounded below. A popular
sufficient condition for this is $V\in L_{2,loc}$ and $V_-\in K_d$, the Kato class
($V_-$ denotes the negative part of $V$). See \cite[Sect.\ 3]{S20}. Conversely,
the fact that $C_0^{\infty}(\R^d)$ functions are in the domain of $H$ implies
that $V\in L_{2,loc}$.

In the discrete case, we assume that $V$ is bounded. It is then of course automatic that
$H$ generates a bounded, self-adjoint operator on $\ell_2(\Z^d)$.

Let us begin with the discrete case. Here the basic lemma is extremely simple.
\begin{Lemma}[finite propagation speed -- discrete case]
\label{Lbasicd}
Consider the operator $H$ from \eqref{sod}.

(a) Let $\varphi_1,\varphi_2\in \ell_2(\Z^d)$, and let $R$ be the distance,
measured in the $|\cdot |_1$ norm on $\Z^d$, between the supports of $\varphi_1$
and $\varphi_2$, respectively. Then
\[
\langle \varphi_1, H^n \varphi_2 \rangle = 0 \quad\text{for}\:\: n=0,1, \ldots , R-1.
\]

(b) Consider two (bounded) potentials $V_1$, $V_2$ and the corresponding
Schr\"odinger operators $H_1$, $H_2$, let $\varphi\in\ell_2(\Z^d)$, and,
similarly to the definition in part (a), let
\[
R = \text{\rm dist} \left( \text{\rm supp }\varphi, \text{\rm supp }(V_1-V_2) \right) .
\]
Then
\[
\langle \varphi, H_1^n \varphi \rangle = \langle \varphi, H_2^n \varphi \rangle
\quad\text{for}\:\: n=0,1,\ldots , 2R .
\]
\end{Lemma}
\begin{proof}
It is clear from the form of $H$ that the support of $H\psi$ is contained
in a $1$-neighborhood (in the $|\cdot |_1$ norm) of the support of $\psi$. Repeatedly applying
this gives (a).

It also follows that in the situation of part (b), $H_1^n\varphi = H_2^n\varphi$ for
$n\le R$ since of course $H_1\psi$ can only be different from $H_2\psi$ if $\psi$ is non-zero at
one of the points where $V_1\not= V_2$. Now it may happen that $(H_1^{R+1}\varphi)(x)\not=
(H_2^{R+1}\varphi)(x)$ for some $x\in\Z^d$, but, according to what has just been observed,
only at points $x$
from the support of $V_1-V_2$, and thus it takes at least $R$ more steps to get back to
the support of $\varphi$ (formally, we could in fact apply part (a) to see this).
\end{proof}

By the spectral theorem, part (b) also says that
\[
\int_{\R} \lambda^n \, d\rho_1(\lambda) =
\int_{\R} \lambda^n \, d\rho_2(\lambda)
\]
for $n=0,1,\ldots , 2R$. Here, $\rho_j$ is the spectral measure for $H_j$ and $\varphi$.
As already discussed in the Introduction, the appropriate continuous substitutes for the powers
$\lambda^n$ seem to be the functions $\cos t\sqrt{\lambda}$,
at least for one-dimensional problems. See also \cite{Remuniv} for further background
information.

This suggests the following continuous analog of Lemma \ref{Lbasicd}.
Since we are assuming that $H$ from \eqref{soc} is bounded below, $\cos t\sqrt{\lambda}$ is
a bounded function on the spectrum of $H$. Also, this function clearly does not depend on a
choice of the square root and is in fact entire in $\lambda$.
\begin{Lemma}[finite propagation speed -- continuous case]
\label{Lbasicc}
Consider the operator $H$ from \eqref{soc}.

(a) Let $\varphi_1,\varphi_2\in L_2(\R^d)$ and define
\[
R = \text{\rm dist} ( \text{\rm supp }\varphi_1, \text{\rm supp }\varphi_2 ) .
\]
Then
\[
\langle \varphi_1 , \cos t\sqrt{H}\, \varphi_2 \rangle = 0 \quad \text{for}\:\: |t|\le R .
\]

(b) Consider two potentials $V_1$, $V_2$, let $\varphi\in L_2(\R^d)$ and define
\[
R = \text{\rm dist} \left( \text{\rm supp }\varphi, \text{\rm supp }(V_1-V_2) \right) .
\]
Then
\[
\langle \varphi, \cos t\sqrt{H_1}\, \varphi \rangle =
\langle \varphi, \cos t\sqrt{H_2}\, \varphi \rangle \quad \text{for}\:\: |t|\le 2R .
\]
\end{Lemma}
\begin{proof}
By a routine approximation argument, we may and will assume that all functions
$V$ and $\varphi$ are in $C_0^{\infty}(\R^d)$. Indeed, since the operators
$\cos t\sqrt{H}$ are bounded, we can certainly approximate the $\varphi$'s in
$L_2(\R^d)$ by smooth
functions, and we then have convergence of the scalar products we are interested in.
We then pick $V_n\in C_0^{\infty}(\R^d)$ so that $H_n=-\Delta+V_n\to H$ in strong
resolvent sense. This is possible by \cite[Theorem VIII.25(a)]{RS1}. But then also
$\cos t\sqrt{H_n} \to \cos t\sqrt{H}$ strongly, by \cite[Theorem VIII.20(b)]{RS1}.
Finally, we can do these approximations in such a way that $\liminf R_n\ge R$ for
the separations $R$ defined in the Lemma.

Now let $u=\cos t\sqrt{H}\varphi_2$. Then $u$ solves
\[
u_{tt} = -Hu, \quad\quad u(0)=\varphi_2,\quad u_t(0)=0 .
\]
Originally, this needs to be interpreted as an equation for functions of $t$ taking
values in $L_2(\R^d)$, but $V$ and $\varphi$ are smooth now, so the regularity results for weak
solutions of (generalized) wave equations actually show that
$u\in C^{\infty}(\R^d\times\R)$ and
\begin{equation}
\label{we}
u_{tt}(x,t)-\Delta u(x,t)=-V(x)u(x,t)
\end{equation}
holds pointwise. Compare \cite[Theorem 7.2.7]{Eva}.

We now need the following classical fact.
\begin{Lemma}[finite propagation speed -- wave equation]
\label{Lwe}
Suppose $u\in C^{\infty}$ solves \eqref{we} and $u(x,0)=u_t(x,0)=0$ for
$|x-x_0|<r$. Then $u\equiv 0$ on $|x-x_0|\le r-|t|$.
\end{Lemma}
\begin{proof}[Proof of Lemma \ref{Lwe}]
We will use an energy estimate.
We can assume that $u$ is real valued and $x_0=0$. Define
\[
E(t) = \int_{B(r-t)} \left( u^2 + u_t^2 + \left| \nabla u \right|^2 \right)\, dx ;
\]
here, $B(q)=\{ x\in\R^d: |x|<q \}$ denotes the ball of radius $q$.
Also, let $S(q)$ be the sphere $S(q)=\{x: |x|=q \}$,
and write $\sigma$ for the surface measure on $S(q)$.
We will now compute $E'$:
\[
E' = - \int_{S(r-t)} \left( u^2+ u_t^2 + \left| \nabla u \right|^2 \right)\, d\sigma
+ 2\int_{B(r-t)} ( uu_t+u_t u_{tt} + \nabla u \cdot \nabla u_t )\, dx
\]
An integration by parts and use of \eqref{we} allow us to write the second integral
in the form
\[
2 \int_{B(r-t)} (1-V)uu_t \, dx + 2 \int_{S(r-t)} u_t\, \textbf{n}\cdot \nabla u \, d\sigma ,
\]
where $\textbf{n}$ is the outer normal unit vector on the sphere. By the Cauchy-Schwarz inequality,
the last term can be estimated by
\[
2 \left( \int_{S(r-t)} u_t^2\, d\sigma \right)^{1/2} \left( \int_{S(r-t)}
\left| \nabla u \right|^2\, d\sigma \right)^{1/2} .
\]
Putting things together, we thus see that
\[
E' \le 2 \int_{B(r-t)}(1+ |V|)|uu_t|\, dx .
\]
Now $V$ is bounded, so a final
application of the Cauchy-Schwarz inequality shows that
\[
E' \le C \left( \int_{B(r-t)} u^2\, dx \right)^{1/2}\left( \int_{B(r-t)} u_t^2\, dx \right)^{1/2}
\le C E .
\]
Since $E(0)=0$, Gronwall's Lemma implies that $E\equiv 0$. This proves the claim (for $t>0$,
but $u(x,-t)$ satisfies the same equation).
\end{proof}
Part (a) of Lemma \ref{Lbasicc} now follows easily: If $x_0$ is an arbitrary point from the support
of $\varphi_1$, then, by assumption, $\varphi_2(x)=0$ for all $|x-x_0|<R$. Lemma \ref{Lwe} now shows that
$(\cos t\sqrt{H} \varphi_2 )(x)=0$ if $|x-x_0|\le R-|t|$. In particular,
$(\cos t\sqrt{H} \varphi_2 )(x_0)=0$ for $|t|\le R$. Since $x_0\in\textrm{supp }\varphi_1$ was
arbitrary, it follows that
$\langle \varphi_1, \cos t\sqrt{H} \varphi_2\rangle = 0$ for these $t$, as claimed.

To prove part (b), let $u_j = \cos t\sqrt{H_j}\varphi$. The argument is similar
to the proof of Lemma \ref{Lbasicd}(b): It takes $T$ units of time to reach points where $V_1\not= V_2$
(starting from the support of $\varphi$), and then another time span $T$ to get back to the support
of $\varphi$. To write this down more formally, first note that Lemma \ref{Lwe} shows that
$u_j(x,t)=0$ whenever $\textrm{dist}(x,\text{supp }\varphi) > |t|$. This shows that for $|t|\le R$,
the difference function $u=u_2-u_1$ solves
\begin{equation}
\label{2.1}
u_{tt} - \Delta u = -V_1 u
\end{equation}
for all $x\in\R^d$. Indeed, if $\textrm{dist}(x,\text{supp }\varphi)<R$, \eqref{2.1} follows from the
fact that then $V_1(x)=V_2(x)$, and if $\textrm{dist}(x,\text{supp }\varphi)> R$, then, as we have
just seen, $u_1(x,t)=u_2(x,t)=0$ for $|t|\le R$.

Since $u(0)=u_t(0)=0$, Lemma \ref{Lwe}, applied to \eqref{2.1}, shows that $u\equiv 0$ for $|t|\le R$.
As argued above, if $\textrm{dist}(x,\text{supp }\varphi)<R$, then \eqref{2.1} in fact holds for
\textit{all} $t$, so a final application of Lemma \ref{Lwe} (to $u(t\pm R)$) shows that $u(t)=0$
on the support of $\varphi$ for $|t|\le 2R$, as desired.
\end{proof}
\section{Decay of kernels}
We begin our discussion of applications of Lemmas \ref{Lbasicd}, \ref{Lbasicc}
with the first problem mentioned in the Introduction, and we first deal with
the discrete case. We will use the following classical result from
approximation theory:
\begin{Theorem}[Jackson]
\label{Tjackson}
Let $f\in C^{\infty}[-1,1]$. Then for every $n\in\N$ and $R\ge n$,
there exist polynomials $p_R$, $\deg p_R \le R$, so that
\[
\| f - p_R \|_{\infty} \le \|f^{(n+1)}\|_1 \left( \frac{5}{R+1} \right)^n .
\]
\end{Theorem}
See \cite[Sect.\ 4.6]{Che} or \cite[Sect.\ 1.1]{Riv} for quick proofs (the second
reference has bigger constants) and \cite{Tim} for a more comprehensive discussion
of these issues, including optimal results.
To obtain the result in the form stated above, use the inequality
\[
N(N-1)\cdots (N-k+1) \ge \left( \frac{N}{e} \right)^k
\]
and note that the modulus of continuity of a function $g$, defined as
$\omega(g,\delta) = \sup_{|x-y|\le\delta} |g(x)-g(y)|$, can obviously be estimated by
$\omega(g,\delta) \le \|g'\|_1$ if $g$ is differentiable.

We'll use the following notation: For $x\in\Z^d$, let $\delta_x$ be
the unit vector located at $x$ (so $\delta_x(x)=1$, $\delta_x(y)=0$ if $y\not= x$).
Moreover, $\sigma(H)$ denotes, as usual, the spectrum of $H$. The estimates we
are about to prove will depend on the diameter of the spectrum.
\begin{Theorem}
\label{T4.1}
Suppose that $\sigma (H)\subset [a,b]$ and
$f\in C^{\infty}[a,b]$. Then, for $n=1,2, \ldots , |x-y|_1 -1$,
\[
\left| \langle \delta_x, f(H) \delta_y \rangle \right| \le
\|f^{(n+1)}\|_{L_1(a,b)} \left( \frac{5(b-a)}{2|x-y|_1} \right)^n .
\]
\end{Theorem}
\begin{proof}
Given Lemma \ref{Lbasicd}, the proof is straightforward: Let
\[
s = \frac{2\lambda - a -b}{b-a},\quad
g(s) = f\left( \frac{a+b}{2} + \frac{b-a}{2}\, s \right) ,
\]
and approximate $g$ on $[-1,1]$ by a polynomial of degree $\le R-1$,
with $R\equiv |x-y|_1$:
\[
g(s) = p_{R-1}(s) + e_{R-1}(s),\quad\quad -1\le s\le 1
\]
By Lemma \ref{Lbasicd}(a),
\[
\left\langle \delta_x, p_{R-1}\left( \frac{2H-a-b}{b-a}\right)
\delta_y\right\rangle = 0 ,
\]
hence
\[
\left| \langle \delta_x, f(H) \delta_y \rangle \right| \le
\max_{-1\le s\le 1} |e_{R-1}(s)| .
\]
By Theorem \ref{Tjackson}, $p_{R-1}$ can be chosen so that
\[
|e_{R-1}(s)| \le \left( \frac{5}{R} \right)^n
\|g^{(n+1)}\|_{L_1(-1,1)} .
\]
Since $\|g^{(n+1)}\|_{L_1(-1,1)} = \left( \frac{b-a}{2} \right)^n \|f^{(n+1)}\|_{L_1(a,b)}$,
the asserted estimate follows.
\end{proof}

The discussion of the continuous case proceeds along similar lines. We will find
it convenient to assume
that $H\ge 0$, where now $H$ is the continuous Schr\"odinger operator from
\eqref{soc}; the general case ($H$ semibounded below) can of course be reduced
to this situation by adding a suitable constant to $H$ (respectively $V$).
\begin{Theorem}
\label{T4.2}
Assume that $H\ge 0$.
Let $f\in C^{\infty}[0,\infty)$, $\varphi_1,\varphi_2\in L_2(\R^d)$, and define
\[
R = \text{\rm dist} ( \text{\rm supp }\varphi_1, \text{\rm supp }\varphi_2 ) .
\]
Also, let $g(k)=f(k^2)$ and assume that $g^{(n)}\in L_1(0,\infty)$ for
all $n\ge 0$. Then, for each $n\in\N$, we have that
\[
\left| \langle \varphi_1, f(H) \varphi_2 \rangle \right| \le
\frac{2\|\varphi_1\|\, \|\varphi_2\|}{\pi n}\, \frac{\|g^{(n+1)}\|_{L_1(0,\infty)}}{R^n} .
\]
\end{Theorem}
\begin{proof}
Expand $f$ in terms of the cosine functions from Lemma \ref{Lbasicc}:
\begin{align*}
f(\lambda) & = \int_0^{\infty} \widetilde{f}(t) \cos t\sqrt{\lambda} \, dt ,\\
\widetilde{f}(t) & = \frac{1}{\pi} \int_0^{\infty} f(\lambda) \cos t\sqrt{\lambda}\,
\frac{d\lambda}{\sqrt{\lambda}} .
\end{align*}
Since $g\in L_1$ or, equivalently, $f\in L_1((0,\infty), d\lambda/\sqrt{\lambda})$,
the formula for $\widetilde{f}$ in fact holds \textit{pointwise.} (The general
theory only guarantees equality in $L_2$ sense.) Making the substitution $k=\sqrt{\lambda}$,
we can write this as
\[
\widetilde{f}(t) = \frac{2}{\pi} \int_0^{\infty} g(k) \cos tk\, dk
= \frac{1}{\pi} \int_{-\infty}^{\infty} g(k) e^{itk}\, dk ,
\]
where we have extended $g$ in the obvious way to all of $\R$. An inductive argument shows
that then $g\in C^{\infty}(\R)$ (the only issue being existence of the derivatives at zero).
We may now integrate by parts $n+1$ times, and since $g^{(j)}\in L_1$ for all $j$, there
are no boundary terms. Thus
\begin{equation}
\label{4.1}
\left| \widetilde{f}(t) \right| \le \frac{1}{\pi t^{n+1}} \|g^{(n+1)}\|_{L_1(\R)} =
\frac{2}{\pi t^{n+1}} \|g^{(n+1)}\|_{L_1(0,\infty)} .
\end{equation}
Now break up $f$ as
\[
f(\lambda) = \int_0^R \widetilde{f}(t) \cos t\sqrt{\lambda}\, dt + \int_R^{\infty}
\widetilde{f}(t) \cos t\sqrt{\lambda}\, dt \equiv h(\lambda) + e(\lambda) .
\]
Again, the rapid decay of $\widetilde{f}$ ensures that everything holds pointwise
for $\lambda\ge 0$. Moreover,
writing $\rho(M)=\langle \varphi_1, E(M) \varphi_2\rangle$
for the spectral measure of $H$ and $\varphi_1$, $\varphi_2$, we have that
\[
\langle \varphi_1, h(H) \varphi_2 \rangle = \int_0^R dt\, \widetilde{f}(t) \int d\rho(\lambda)
\cos t\sqrt{\lambda} = 0,
\]
by Fubini's Theorem and Lemma \ref{Lbasicc}(a). Therefore,
\[
\left| \langle \varphi_1, f(H) \varphi_2 \rangle \right| \le \|\varphi_1\| \, \|\varphi_2\|\,
\sup_{\lambda\ge 0} \left| e(\lambda) \right| .
\]
But the definition of the error term $e$ and \eqref{4.1} imply that
\[
\left| e(\lambda ) \right| \le \int_R^{\infty} \left| \widetilde{f}(t) \right| \, dt
\le \frac{2}{\pi n R^n} \|g^{(n+1)}\|_{L_1(0,\infty)} ,
\]
and the proof is complete.
\end{proof}

As an aside, note the slightly different approach to the approximation problem in this
proof, as opposed to the proof of Theorem \ref{T4.1}: We actually used
an $L_2$ approximation, which, however, also is a very reasonable uniform
approximation for smooth functions. A similar device could have been used above
to replace Theorem \ref{Tjackson}: We can expand $f(x)=\sum a_n T_n(x)$ in
terms of Chebyshev polynomials, and then an integration by parts argument
lets us bound the $a_n$. The final result of this is a bound of the form
$\|f-p_R\|_{\infty} \lesssim R^{-n} \|g^{(n+1)}\|_1$, with $g(\theta)=f(\cos\theta)$.

Returning to the statements of Theorems \ref{T4.1}, \ref{T4.2}, we remark that
if one places some restrictions on the growth of the
$L_1$ norms of the derivatives, one can minimize the
bounds established above over $n$ and obtain
new estimates. We will pursue this theme in Sect.\ 5, where we will
also compare Theorem \ref{T4.2} to the closely related work of
Bouclet, Germinet, and Klein \cite{BGK, GK}. These references use
the Helffer-Sj\"ostrand formula
and a Combes-Thomas estimate as their main tools.
\section{A priori estimates on spectral measures}
In a similar way, part (b) of Lemmas \ref{Lbasicd} and \ref{Lbasicc},
respectively, yields results addressing the second problem mentioned above.
This section is inspired by recent work of Germinet, Kiselev,
and Tcheremchantsev \cite[Lemma A.1]{GKT}. We will take a look at this from the point
of view suggested by the material of Sect.\ 2. The treatment of \cite{GKT}
is again based on the Helffer-Sj\"ostrand formula
and a Combes-Thomas estimate. In the one-dimensional case, I earlier discussed related
problems in \cite{Remuniv} from a point of view remotely reminiscent of the one taken here.

As in the previous section, we begin with the discrete case.
\begin{Theorem}
\label{T3.3}
Let $f\in C_0^{\infty}(-1,1)$, $\lambda_0\in\R$, and define, for $0<\epsilon\le 1$,
\[
f_{\epsilon}(\lambda) = f \left( \frac{\lambda-\lambda_0}{\epsilon} \right).
\]
Let $\varphi\in \ell_2(\Z^d)$, and let $V_1$, $V_2$ be two bounded potentials that
agree on an $R$-neighborhood (with respect to the $|\cdot |_1$ norm on $\Z^d$)
of the support of $\varphi$. Determine an interval $[a,b]\supset \sigma (H_1)
\cup \sigma (H_2)$.
Then, for $n=1,2,\ldots, 2R$,
\[
\left| \langle \varphi, f_{\epsilon}(H_1) \varphi\rangle -
\langle \varphi, f_{\epsilon}(H_2) \varphi \rangle\right| \le
2 \|\varphi\|^2 \|f^{(n+1)}\|_1 \left( \frac{5 (b-a)}{4\epsilon R} \right)^n .
\]
\end{Theorem}
Here, we have tacitly extended the function $f$ to all of $\R$ by setting $f=0$ outside $(-1,1)$.

For an interesting application,
let $f$ approximate the characteristic function of an interval. Since
$\langle \varphi, f_{\epsilon}(H_j) \varphi \rangle = \int f_{\epsilon}(\lambda)
\, d\rho_j(\lambda)$, where $\rho_j$ is the spectral measure of $H_j$ and $\varphi$,
Theorem \ref{T3.3} now says that it is possible to approximately compute $\rho(I)$
for an interval $I\subset\R$ with high accuracy, provided $V$ is known on an $R$-neighborhood
of $\textrm{supp }\varphi$ and $R|I|\gg 1$.
See \cite[Sect.\ 8]{Kis} for such an application of \cite[Lemma A.1]{GKT}.

\begin{proof}
Let again
\[
s = \frac{2\lambda - a -b}{b-a},\quad
g(s) = f_{\epsilon}\left( \frac{a+b}{2} + \frac{b-a}{2}\, s \right) ,
\]
and write, according to Theorem \ref{Tjackson},
\[
g(s) = p_{2R}(s) + e_{2R}(s), \quad\quad  -1\le s\le 1,
\]
with a polynomial $p_{2R}$ of degree $\le 2R$. By Lemma \ref{Lbasicd}(b),
\[
\left\langle \varphi, p_{2R}\left( \frac{2H_1-a-b}{b-a} \right) \varphi\right\rangle
= \left\langle \varphi, p_{2R}\left( \frac{2H_2-a-b}{b-a} \right) \varphi\right\rangle
\]
To bound the error $e_{2R}$, note that
\[
\frac{d^ng}{ds^n} =
\left( \frac{b-a}{2\epsilon} \right)^n
f^{(n)}\left( \frac{b-a}{2\epsilon}\, s + \frac{a+b}{2\epsilon} - \frac{\lambda_0}{\epsilon} \right) ,
\]
thus
\begin{align*}
\|g^{(n+1)}\|_1 & = \left( \frac{b-a}{2\epsilon} \right)^{n+1} \int_{-1}^1
\left| f^{(n+1)}\left( \frac{b-a}{2\epsilon}\, s + \frac{a+b}{2\epsilon} - \frac{\lambda_0}{\epsilon} \right) \right|
\, ds \\
& = \left( \frac{b-a}{2\epsilon} \right)^n \int_{(a-\lambda_0)/\epsilon}^{(b-\lambda_0)/\epsilon}
\left| f^{(n+1)}(t)\right| \, dt
\le \left( \frac{b-a}{2\epsilon} \right)^n \|f^{(n+1)}\|_1 ,
\end{align*}
and now Theorem \ref{Tjackson} shows that for $-1\le s\le 1$,
\[
|e_{2R}(s)| \le \left( \frac{5(b-a)}{4\epsilon R} \right)^n
\|f^{(n+1)}\|_1 .
\]
This completes the proof.
\end{proof}

Let us now discuss the continuous case. As in the previous section,
we will assume that $H_j\ge 0$, which, of course, is just a normalization.
Again, the more natural variable is
$k=\sqrt{\lambda}$ and thus the first version of our basic result reads as follows:
\begin{Theorem}
\label{T3.2}
Let $f\in C_0^{\infty}(-1,1)$, $\lambda_0>0$, and define, for $0<\epsilon\le\sqrt{\lambda_0}$,
\[
f_{\epsilon}(\lambda) = f \left( \frac{\sqrt{\lambda}-\sqrt{\lambda_0}}{\epsilon} \right) .
\]
Let $\varphi\in L_2(\R^d)$ and
suppose that $V_1$, $V_2$ agree on an $R$-neighborhood of the support of $\varphi$. Then,
for every $n\in\N$,
\[
\left| \langle \varphi , f_{\epsilon}(H_1) \varphi \rangle
- \langle \varphi , f_{\epsilon}(H_2) \varphi \rangle \right|
\le \frac{4\|\varphi\|^2 \|f^{(n+1)}\|_1}{n\pi(2\epsilon R)^n} .
\]
\end{Theorem}
\begin{proof}
The basic ideas are familiar by now, so it will suffice to provide a sketch
of the argument. Expand $f_{\epsilon}$:
\begin{align*}
f_{\epsilon}(\lambda) & = \int_0^{\infty} \widetilde{f}_{\epsilon}(t) \cos t\sqrt{\lambda}\, dt ,\\
\widetilde{f}_{\epsilon}(t) & = \frac{1}{\pi} \int_0^{\infty} f_{\epsilon}(\lambda)
\cos t\sqrt{\lambda} \, \frac{d\lambda}{\sqrt{\lambda}}
\end{align*}
Use $s=(\sqrt{\lambda}-
\sqrt{\lambda_0})/\epsilon$ as the variable in the integral defining $\widetilde{f}_{\epsilon}$:
\[
\widetilde{f}_{\epsilon}(t) = \frac{2\epsilon}{\pi}
\int_{-\infty}^{\infty} f(s) \cos\left( \epsilon ts+t\sqrt{\lambda_0}\right) \, ds
\]
Integrating by parts, we thus see that
\begin{equation}
\label{3.5}
|\widetilde{f}_{\epsilon}(t)| \le \frac{2\epsilon\|f^{(n+1)}\|_1}{\pi(\epsilon t)^{n+1}} .
\end{equation}
We can now argue as in the last part of the proof of Theorem \ref{T4.2} to finish the proof.
More specifically, write
\[
f_{\epsilon}(\lambda) = \int_0^{2R} \widetilde{f}_{\epsilon}(t) \cos t\sqrt{\lambda}\, dt
+ \int_{2R}^{\infty} \widetilde{f}_{\epsilon}(t) \cos t\sqrt{\lambda}\, dt \equiv g + e,
\]
note that $\langle \varphi, (g(H_1)-g(H_2)) \varphi \rangle =0$ by Lemma \ref{Lbasicc}(b),
and use \eqref{3.5} to bound the error term $e$.
\end{proof}

It is of course also possible to use the original variable $\lambda$, although things
are slightly less elegant then. We continue to assume that $H_j\ge 0$.
\begin{Theorem}
\label{T3.1}
Let $f\in C_0^{\infty}(-1,1)$, $\lambda_0>0$, and define, for $0<\epsilon\le \min\{
1, \lambda_0/2 \}$,
\[
f_{\epsilon}(\lambda) = f \left( \frac{\lambda-\lambda_0}{\epsilon} \right).
\]
Let $\varphi\in L_2(\R^d)$, and let $V_1$, $V_2$ be two potentials that
agree on an $R$-neighborhood of the support of $\varphi$. Then, for all $n\in\N$,
\begin{equation}
\label{2.3}
\left| \langle \varphi, f_{\epsilon}(H_1)\varphi\rangle
- \langle \varphi, f_{\epsilon}(H_2)\varphi\rangle \right|
\le \|\varphi\|^2\, \frac{C_n}{(\epsilon R)^n} .
\end{equation}
The constant $C_n$ depends on $\|f^{(j)}\|_1$ for $j=1,2,\ldots, n+1$ and $\lambda_0$,
but is independent of $\epsilon, R, V_1, V_2, \varphi$.
\end{Theorem}
\begin{proof}
We now use the variable $s=(\lambda-\lambda_0)/\epsilon$ in the expansion
\begin{align*}
f_{\epsilon}(\lambda) & = \int_0^{\infty} \widetilde{f}_{\epsilon}(t)
\cos t\sqrt{\lambda}\, dt , \\
\widetilde{f}_{\epsilon}(t) & = \frac{1}{\pi} \int_0^{\infty} f_{\epsilon}(\lambda)
\cos t\sqrt{\lambda}\, \frac{d\lambda}{\sqrt{\lambda}} .
\end{align*}
So
\[
\widetilde{f}_{\epsilon}(t) = \frac{\epsilon}{\pi} \int_{-\infty}^{\infty}
f(s) \cos t\sqrt{\lambda_0+\epsilon s} \, \frac{ds}{\sqrt{\lambda_0+\epsilon s}}
\]
The integral is still oscillatory and thus should
be small for large $t$. To make this precise, write $\sqrt{\lambda_0+x}\equiv\omega (x)$, note
that $1/\omega = 2\omega'$ and consider
\begin{equation}
\label{2.2}
\int_{-\infty}^{\infty} f(s)\omega'(\epsilon s) e^{it\omega(\epsilon s)}\, ds .
\end{equation}
We introduce the differential expression
\[
D = \frac{-i}{\epsilon t \omega'(\epsilon s)}\, \frac{d}{ds} ,
\]
so that $D e^{it\omega (\epsilon s)} = e^{it\omega (\epsilon s)}$. Thus the integral from
\eqref{2.2} equals
\[
\int_{-\infty}^{\infty} f(s)\omega'(\epsilon s) D^n e^{it\omega (\epsilon s)}\, ds =
\int_{-\infty}^{\infty} e^{it\omega (\epsilon s)}D'^n f(s)\omega'(\epsilon s)\, ds,
\]
where
\[
D' = \frac{d}{ds} \, \frac{i}{\epsilon t \omega'(\epsilon s)}
\]
is the transpose of $D$. We can evaluate $D'^n f\omega' = i(\epsilon t)^{-1} D'^{n-1}f'$, using the product rule.
We obtain a sum of many terms, each of which is of the form
\[
c (\epsilon t)^{-n}
\epsilon^{n-k} \frac{\omega^{(p_1)} \cdots \omega^{(p_m)}}{(\omega')^P}\, f^{(k)} ,
\]
with $1\le k\le n$, $m\ge 0$, $p_j\ge 2$ and $P\ge n$; the argument of the derivatives
of $\omega$ is $\epsilon s$. It follows that $|\eqref{2.2}| \lesssim (\epsilon t)^{-n}$,
with a constant that depends on $n$, $\lambda_0$ and bounds on the derivatives of $f$.

The rest of the argument proceeds as in the previous proof.
\end{proof}
\section{Some remarks on Gevrey type functions}
Usually, one defines the class of Gevrey functions on an interval $I\subset\R$ as
follows (we restrict to the one-dimensional case right away because that is all we will
need here): By definition, $f\in G_s(I)$ if $f\in C^{\infty}(I)$ and
\begin{equation}
\label{5.2}
|f^{(n)}(x)| \le C^{n+1} n^{sn} \quad\quad (x\in K, n\in\N_0 )
\end{equation}
for every compact subset $K\subset I$. Here, $s\ge 1$, and the constant $C$ may
depend on $f$ and $K$.
The functions $f\in G_1(I)$ are in fact real analytic (see \cite[Theorem 19.9]{Rud}).
On the other hand, if $s>1$, then $G_s(I)$
contains compactly supported functions (easily constructed with the help
of the functions $\exp(-x^{-b})$ for suitable $b>0$).

For our purposes, it is clear from the results discussed in Sect.\ 3, 4 that \textit{global}
control on the derivatives is more relevant. This was also (and previously)
recognized in \cite{BGK}. The following definition seems most appropriate:
\begin{Definition}
\label{D5.1}
Let $I\subset\R$ be an interval and let $s\ge 1$.
We say that $f\in G_s^1(I)$ if $f\in C^{\infty}(I)$ and there exists a constant $C>0$ so that
\[
\| f^{(n)}\|_{L_1(I)} \le C^{n+1} n^{sn} \quad\quad \textrm{for all }n\in\N_0 .
\]
\end{Definition}
The relation of this to $G_s$ can be clarified by making the following quick observations:
Suppose that $f\in G_s^1(I)$. Then there exists a constant $A>0$ (any positive $A$ with
$A|I|>1$ will do), so that for every $n\in\N_0$, we can find an $x_0^{(n)}\in I$ so that
$\left| f^{(n)}(x_0^{(n)}) \right| \le A C^{n+1} n^{sn}$.
But then for arbitrary $x\in I$, we have that
\begin{align*}
\left| f^{(n)}(x) \right| & \le \left| f^{(n)}(x_0^{(n)}) \right| + \| f^{(n+1)}\|_{L_1(I)}
\le AC^{n+1} n^{sn} + C^{n+2} (n+1)^{s(n+1)}\\
& = C^{n+1} n^{sn} \left( A
+ C(n+1)^s \left( 1 + \frac{1}{n} \right)^{sn} \right) .
\end{align*}
It follows that there exists a new constant $D$, independent of $n$, so that
$\left| f^{(n)}(x) \right| \le D^{n+1}n^{sn}$ on $I$. In particular, $f\in G_s(I)$,
but we have in fact obtained the stronger statement that the constants $C$ from
\eqref{5.2} can be taken to be independent of $K$. We denote the set of functions
satisfying such a uniform Gevrey condition by $G_s^{\text{\rm unif}}(I)$. Also,
it is obvious, by integrating the pointwise bounds, that $G_s(K) \subset G_s^1(K)$
for \textit{compact} intervals $K\subset\R$. We have thus proved the following.
\begin{Proposition}
\label{P5.1}
(a) $G_s^1(I) \subset G_s^{\text{\rm unif}}(I)\subset G_s(I)$;\\
(b) $G_s^1(K) = G_s(K)$ if $K\subset\R$ is a compact interval.
\end{Proposition}
If our functions are in $G_s^1$, we can obtain more explicit information from
the results of the preceding sections. We illustrate this
with the Gevrey versions of Theorems \ref{T4.1} and \ref{T4.2}, respectively.
Of course, there are similar Gevreyzations of Theorems \ref{T3.3} and \ref{T3.2},
which we won't make explicit.
\begin{Corollary}
\label{C5.1}
In the situation of Theorem \ref{T4.1}, suppose that $f\in G_s^1[a,b]$. Then there exist
constants $C,\gamma >0$ so that
\[
| \langle \delta_x, f(H) \delta_y \rangle | \le C \exp\left( -\gamma |x-y|_1^{1/s}\right) .
\]
\end{Corollary}
\begin{proof}
It suffices to take $n\approx (R/C)^{1/s}$ with a suitable constant $C$ in
Theorem \ref{T4.1}. To spell this out more explicitly, first note that
by Theorem \ref{T4.1} and the definition of $G_s^1$, we have that
\begin{equation}
\label{5.1}
| \langle \delta_x, f(H) \delta_y \rangle | \le \left( \frac{Bn^s}{R} \right)^n
\quad\quad (n\in\N,\: R\equiv |x-y|_1 > n)
\end{equation}
for a suitable constant $B>0$. We may assume that
$B\ge 2/e$, and we then pick $n$ so that
\[
\frac{R}{2eB} \le n^s \le \frac{R}{eB} ,
\]
provided there actually exists an integer satisfying these bounds.
This, however, will certainly be the case if $R$ is large enough.
Since $s\ge 1$, $n$ is then not larger than $R-1$, and thus \eqref{5.1} shows that
\[
| \langle \delta_x, f(H) \delta_y \rangle | \le e^{-n} \le e^{-\gamma R^{1/s}} .
\]
This proves the asserted bound for large $R$, and validity for all $R$ is then achieved by
simply adjusting the constant $C$.
\end{proof}
As observed above in Proposition \ref{P5.1}(b), $G_s^1[a,b] = G_s[a,b]$, so we don't
really need Definition \ref{D5.1} here. The class $G_s^1$ does become relevant, however, in the continuous
case because then the spectra are unbounded and global bounds are needed.
We have the following analog of Corollary \ref{C5.1}.
\begin{Corollary}
\label{C5.2}
In the situation of Theorem \ref{T4.2}, suppose that $g\in G_s^1([0,\infty))$.
Then there exist
constants $C,\gamma >0$ so that
\[
| \langle \varphi_1, f(H) \varphi_2 \rangle | \le C \|\varphi_1 \|\,
\|\varphi_2\| \exp\left( -\gamma R^{1/s}\right) .
\]
\end{Corollary}

\noindent
The \textit{proof} is completely analogous to the proof of Corollary \ref{C5.1}.\hfill$\square$

\medskip
In \cite{BGK}, Bouclet, Germinet, and Klein introduce the class of
$C^{\infty}(I)$ functions $f$ that obey estimates of the form
\[
\int_I (1+|\lambda |)^{n-1} \left| f^{(n)}(\lambda)\right| \, d\lambda \le C^{n+1} n^{sn} ,
\]
and they go on to prove that if this holds on an open interval $I$ containing the
spectrum of $H$, then
\[
| \langle \varphi_1, f(H) \varphi_2 \rangle | \lesssim \exp\left( -\gamma R^{(1/s)-\epsilon}\right)
\]
for every $\epsilon>0$ \cite[Theorem 1.4]{BGK}.

It may therefore be interesting to relate our hypothesis that $g\in G_s^1$, where $g(k)=f(k^2)$,
to such a condition.
\begin{Theorem}
\label{T5.1}
Let $s\ge 1$. Suppose that
$f\in C^{\infty}[0,\infty)$ and
\[
\int_0^{\infty} (1+\lambda)^{(n-1)/2} \left| f^{(n)}(\lambda) \right| \, d\lambda \le
C^{n+1} n^{sn} \quad\quad (n\in\N_0) .
\]
Then $g(k)=f(k^2)\in G_s^1([0,\infty))$.
\end{Theorem}
\begin{proof}
This is elementary but a bit tedious.
Working out the derivatives with the help of the chain and product rules, we see that
$g^{(n)}$ is of the form
\[
g^{(n)}(k) = \sum a_{ij}(n) k^i f^{(j)}(k^2) ,
\]
where the sum ranges over all indices $n/2 \le j\le n$, $0\le i\le j$
satisfying $i=2j-n$. To confirm that this relation between $i$ and $j$ must hold,
one can argue as follows: To produce a contribution of the form $k^i f^{(j)}$,
we clearly must let precisely $j$ of the $n$ derivatives act on $f$. By the chain rule, this gives
$j$ factors of $k$, and the remaining $n-j$ derivatives must then act on these.
As a result, the exponent $j$ decreases by $n-j$; in other words, $i=2j-n$. The other
restrictions on $i, j$ follow from this.

Our task is to extract some information
on the coefficients $a_{ij}(n)$. By taking the derivative in the above representation,
it follows that the $a_{ij}(n)$ obey the recursion
\begin{equation}
\label{5.5}
a_{ij}(n) = (i+1) a_{i+1,j}(n-1) + 2 a_{i-1,j-1}(n-1) .
\end{equation}
It will be convenient to use the difference $d=j-i$ as the parameter indexing the coefficients
$a$. Note that for fixed $n$, we have that $0\le d \le n/2$.
Moreover, $a_{ij}(n)$ can be different from zero
only if $i=n-2d$, $j=n-d$. We'll use the abbreviation
$C_d(n) = a_{n-2d,n-d}(n)$, and we then claim that
\begin{equation}
\label{ind}
0\le C_d(n) \le \frac{2^{n-2d} n^{2d}}{d!} .
\end{equation}
Of course, only the upper bound needs proof, and we can use
induction on $d$. The case $d=0$ corresponds to letting all the
derivatives act on $f$, so $C_0(n)=2^n$. Now assume that \eqref{ind} holds for $d-1$,
with $d\ge 1$. By \eqref{5.5} and the definition of $C_d(n)$,
\[
C_d(n) = (n-2d+1) C_{d-1}(n-1) + 2C_d(n-1) .
\]
We will now iterate this, using the induction hypothesis and the fact that $d\ge 1$
to estimate the first term on the right-hand side. In the first step, we obtain
\begin{align*}
C_d(n) & \le \frac{2^{n-2d+1}}{(d-1)!} (n-1)^{2d-1} + 2C_d(n-1) \\
& = \frac{2^{n-2d+1}}{(d-1)!} (n-1)^{2d-1} + 2((n-1)-2d+1)C_{d-1}(n-2) + 2^2 C_d(n-2) \\
& \le \frac{2^{n-2d+1}}{(d-1)!} \left( (n-1)^{2d-1} + (n-2)^{2d-1} \right) + 2^2 C_d(n-2) .
\end{align*}
Continuing in this way and recalling that $C_d(n)=0$ as soon as $n<2d$, we see that
\[
C_d(n) \le \frac{2^{n-2d+1}}{(d-1)!} \sum_{j=1}^{n-1} j^{2d-1}\le
\frac{2^{n-2d+1}}{(d-1)!} \int_1^n x^{2d-1}\, dx < \frac{2^{n-2d} n^{2d}}{d!} ,
\]
as required.

We are now ready to estimate the integrals $\int_0^{\infty} |g^{(n)}|\, dk$. To prove
Theorem \ref{T5.1}, it clearly suffices to show that
\[
C_d(n) \int_0^{\infty} k^{n-2d} \left| f^{(n-d)}(k^2)\right| \, dk \le C^{n+1} n^{sn}
\]
for $0\le d\le n/2$, with a constant $C$ independent of $d$ and $n$.
We will split these integrals
into two parts, corresponding to $0\le k\le 1$ and $k>1$, respectively.
Consider first $C_d(n) \int_1^{\infty} \cdots$. Using $\lambda=k^2$ as the
variable, we can write this as
\[
\frac{1}{2}\, C_d(n) \int_1^{\infty} \lambda^{(n-2d-1)/2} \left| f^{(n-d)}(\lambda)\right| \, d\lambda,
\]
and, by hypothesis, we have a bound of the form $C_d(n) C^{n+1} (n-d)^{s(n-d)}$.
So, taking \eqref{ind} into account,
we must now show that there exists a constant $C>0$ so that
\[
\frac{n^{2d}}{d^d}\, (n-d)^{s(n-d)} \le C^{n+1} n^{sn}
\]
for all $0\le d\le n/2$. This can be done quite easily, by simply further estimating
$(n-d)^{s(n-d)} \le n^{s(n-d)}$.

A similar argument lets us bound $C_d(n) \int_0^1 \cdots$. We can in fact work with
pointwise bounds. Indeed, by hypothesis and Proposition \ref{P5.1}(b), $f\in G_s[0,1]$,
so, reasoning along the above lines, we see that $g\in G_s[0,1]$, too.
\end{proof}

\end{document}